# SISMIK for brain MRI: Deep-learning-based motion estimation and model-based motion correction in k-space

Oscar Dabrowski, Jean-Luc Falcone, Antoine Klauser, Julien Songeon, Michel Kocher, Bastien Chopard, François Lazeyras, and Sébastien Courvoisier

**Abstract**—MRI, a widespread non-invasive medical imaging modality, is highly sensitive to patient motion. Despite many attempts over the years, motion correction remains a difficult problem and there is no general method applicable to all situations. We propose a retrospective method for motion estimation and correction to tackle the problem of in-plane rigid-body motion, apt for classical 2D Spin-Echo scans of the brain, which are regularly used in clinical practice. Due to the sequential acquisition of k-space, motion artifacts are well localized. The method leverages the power of deep neural networks to estimate motion parameters in k-space and uses a model-based approach to restore degraded images to avoid "hallucinations". Notable advantages are its ability to estimate motion occurring in high spatial frequencies without the need of a motion-free reference. The proposed method operates on the whole k-space dynamic range and is moderately affected by the lower SNR of higher harmonics. As a proof of concept, we provide models trained using supervised learning on 600k motion simulations based on motion-free scans of 43 different subjects. Generalization performance was tested with simulations as well as in-vivo. Qualitative and quantitative evaluations are presented for motion parameter estimations and image reconstruction. Experimental results show that our approach is able to obtain good generalization performance on simulated data and in-vivo acquisitions.

Manuscript submitted for review August 28, 2023. We acknowledge access to the facilities and expertise of the CIBM Center for Biomedical Imaging founded and supported by Lausanne University Hospital (CHUV), University of Lausanne (UNIL), École polytechnique fédérale de Lausanne (EPFL), University of Geneva (UNIGE) and Geneva University Hospitals (HUG).

Oscar Dabrowski is with the University of Geneva, faculty of science, route de Drize 7 1227 Carouge, Switzerland and the Department of Radiology and Medical Informatics, University of Geneva, Switzerland (e-mail: oscar.dabrowski@unige.ch).

Sébastien Courvoisier is with the Department of Radiology and Medical Informatics, University of Geneva, Switzerland and the CIBM Center for Biomedical Imaging, Geneva, Switzerland.

Jean-Luc Falcone is with the Computer Science Department, University of Geneva.

Antoine Klauser is with the Department of Radiology and Medical Informatics, University of Geneva, Switzerland and the CIBM Center for Biomedical Imaging, Geneva, Switzerland.

Julien Songeon is with the Department of Radiology and Medical Informatics, University of Geneva, Switzerland and the CIBM Center for Biomedical Imaging, Geneva, Switzerland.

Michel Kocher is with the EPFL Biomedical Imaging Group (BIG).

Bastien Chopard is with the Computer Science Department, University of Geneva.

François Lazeyras is with the Department of Radiology and Medical Informatics, University of Geneva, Switzerland and the CIBM Center for Biomedical Imaging, Geneva, Switzerland.

We provide a Python implementation at https://gitlab.unige.ch/Oscar.Dabrowski/sismik_mri/.

## I. INTRODUCTION

MRI is an essential imaging modality in medicine, which suffers unfortunately from a great sensitivity to movements that can deteriorate image quality. Motion occurs sequentially in time and is reflected into the sequential k-space acquisition in a mostly predictable manner (see subsection II-A). We make the hypothesis that motion parameters can be retrieved from k-space, given knowledge of the acquisition scheme. The latter can be very diverse in MRI, which makes the problem very complex[1].

Deep learning (DL) - a branch of machine learning (ML) that uses deep artificial neural networks (DNNs) - is attractive due to its ability to deliver fast solutions for highly complex problems, which might not be practically solvable by conventional methods.

Many solutions, operating in image space and using deep convolutional neural networks for "image-to-image" motion correction, have been proposed[2–7]. Although they can produce high quality results, they suffer from reconstruction instabilities[8] such as introducing hallucinations (spurious structures that might be misleading), which is a very significant concern in the medical domain[9–11]. To avoid this caveat, other approaches tackle the problem by detecting motion in image space and use model-based approaches for motion parameter estimation and reconstruction, thereby avoiding the risk of hallucinations[12]. Polak *et al.* used additional low-resolution scout scans as a motion-free reference to guide the search for motion parameters[13].

Another alternative is to work in k-space rather than image space, due to the locality of motion (MRI acquisition is an inherently sequential process) and also to reduce the dependence to specific contrasts or image structures present in the training datasets. Applying deep learning directly in k-space is however inherently more difficult, due to its high dynamic range, as opposed to image space. It is also not clear how to properly normalize k-space to make it suitable for deep learning. Regarding approaches operating in k-space rather than image space, recent work by Eichhorn *et al.* proposed a DL-based approach to classify motion-corrupted k-space lines with performance assessment performed only on simulated datasets[14]. A preliminary investigation to estimate motion from adjacent k-space phase lines, incorporating a model-based correction method was proposed in [15]. The feasibility of this approach was mostly assessed using synthetic rectangle images for which an analytical expression of the Fourier transform is known. Research by Mendes *et al.* [16] have shown that a modified correlation operation applied on adjacent phase lines allows to estimate translational motion in turbo Spin-Echo acquisitions, (provided that a sufficient turbo factor is used). Hossbach *et al.* [17] used a deep neural network to quantify rigid-body motion in k-space. However, they used a motion-free reference to estimate motion parameters and only considered the central (64x64) region of k-space.

Other methods, not based on ML/DL, known as "Autofocusing" - in analogy to optical systems where a sharp image is obtained by adjusting the position of a lens - use a mathematical formalism (usually formulated as a matrix inversion problem) together with the optimization of an image quality metric (typically similar to Shannon entropy) to estimate motion parameters allowing a reduction of motion artifacts[18–21].

It is noteworthy, that the aforementioned approaches are *retrospective*, i.e., perform motion correction after data has been acquired. Another class of methods, known as *prospective*, adapt gradient waveforms in real time to compensate for subject motion during scanning [22].

To broaden the perspective, we should mention work that tackle the problem of non-rigid motion, although this does not apply to bulk head motion, which is rigid. Huttinga *et al.* proposed MR-MOTUS, a model-based approach to estimate 3D non-rigid motion fields [23]. The approach utilizes k-space data and assumes the availability of an artefact-free reference image to perform motion estimation. A deep learning model called LAPNet was also proposed to tackle the problem of non-rigid motion directly in k-space. A deformation field was estimated in k-space based on supervised learned filters for a pair-wise registration of local translational offsets (image domain) corresponding to phase differences (k-space domain) [24]. We should also mention that substantial work has been performed in the domain of non-rigid motion correction for cardiac MRI, using deep learning approaches [25–27].

We propose a novel retrospective deep learning-based approach that overcomes the need for a motion-free reference and is able to learn in a wide range of spatial frequency regions of k-space. Furthermore, any risk of hallucination is avoided *by construction*, i.e., the DNN model only estimates motion parameters and reconstruction is performed leveraging the Fourier theorems using these estimates. Note that approaches referred to as physics-informed neural networks (PINNs) are able to incorporate prior knowledge of physical laws in their architecture and/or associated cost functions, which can act as some kind of regularization to restrict their solution space (as opposed to purely data-driven approaches) [28, 29]. While PINNs have been used to solve problems based on partial differential equations, this is not the approach we have chosen. Our image reconstruction method is rather similar to those used in [17] or [21].

A novel k-space quality metric is also proposed. The idea stems from a 1995 publication, where Wood *et al.*[30] leveraged visible discontinuities in k-space to detect the occurrences of motion. Using the ESPIRiT multi-coil reconstruction approach[31], and a very small neural network (272 trainable parameters), we are able to locate corrupted phase encoding lines and provide a quality score. Moreover, the metric is reference-less. To the best of our knowledge, no *k-space* quality metric has yet been proposed. Conventional reference-less quality metrics operate in image space [32–36]. A diagram describing the proposed motion-correction pipeline is provided in Figure 1. It is not intended to cover all possible trajectories in k-space, rather we focus on 2D Spin-Echo sequences [37] with Cartesian sampling to demonstrate the ability of DNNs to retrieve motion parameters in k-space.

The following sections are organized as follows: Section II describes the physical model and the methodology. This entails the design of the k-space quality metric, motion estimation with the proposed DNN, a detailed description of its architecture along with the training procedure. Then, the proposed model-based motion correction method is presented and the section is concluded by a description of the experiments performed. Experimental results, including a comparison with another approach, are presented in Section III and discussed in Section IV.

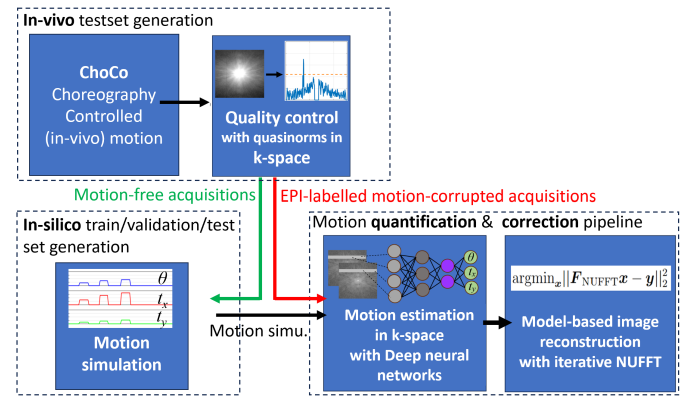

Fig. 1: Visualization of the motion acquisition, simulation, quantification and correction pipeline. The quality control metric plays an important role in the development/verification process, in particular, to verify in-vivo motion-corrupted acquisition as well as simulation consistency.

## II. METHODS

This section starts with a mathematical description of the physical process involved in the MRI acquisitions considered in this paper and a description of the rigid-body motion model. Subsequently, we describe a novel k-space quality metric, followed by a motion simulation procedure, used to generate large datasets for training deep learning models for motion estimation, which we then proceed to present. We describe the proposed DNN architecture and the model-based motion correction procedure, and conclude the section with a description of the experiments performed.

*A. Description of the problem*

The physical process that takes place for each phase encoding line of a classical Spin-Echo acquisition - which we used for our experiments - can be described by the following equation:

$$S_c(k_x, k_y) = \int \int C_c(x,y)\rho(x,y)e^{-i2\pi x k_x}e^{-i2\pi y k_y}\,dx\,dy \quad (1)$$

Where the integrals run over the whole field of view, $\rho \in \mathbb{R}$ is the ($T_1$ or $T_2$) weighted spin density, $C_c \in \mathbb{C}$ is the $c$-th coil sensitivity profile, $i^2 = -1$ and $k_x \in \mathbb{R}$, $k_y \in \mathbb{R}$ are the wavenumbers in the frequency and phase encoding directions, respectively.

During a classical Spin-Echo acquisition, phase encoding allows the acquisition of a single line of k-space at each TR (in the order of seconds). Frequency encoding is much faster (in the order of milliseconds), therefore, only motion in the phase encoding direction is considered.

When a patient moves during the acquisition of a specific phase encoding line ($k_y$), the position-dependent resonance frequency of spins induced by magnetic field gradients introduces a displacement of the acquired points in k-space. This phenomenon typically manifests as artifacts in image space, characterized by the appearance of "ringing" or "ghosting" effects. This mismapping can be interpreted as sampling the wrong spatial frequencies, while the patient is static (as opposed to the patient moving while sampling the regular Fourier coefficients). This perspective conceptually leads to the usage of the non-uniform fast Fourier transform (NUFFT)[38–41] for reconstructing motion-corrupted scans (process described in more details in subsection II-E).

Rigid-body motion can be described by the following homogeneous matrix:

$$\begin{pmatrix} 1 & 0 & \tau_x \\ 0 & 1 & \tau_y \\ 0 & 0 & 1 \end{pmatrix} \begin{pmatrix} \cos(\theta) & -\sin(\theta) & 0 \\ \sin(\theta) & \cos(\theta) & 0 \\ 0 & 0 & 1 \end{pmatrix} \begin{pmatrix} 1 & 0 & -\tau_x \\ 0 & 1 & -\tau_y \\ 0 & 0 & 1 \end{pmatrix}$$
$$= \begin{pmatrix} \cos(\theta) & -\sin(\theta) & -\tau_x\cos(\theta) + \tau_y\sin(\theta) + \tau_x \\ \sin(\theta) & \cos(\theta) & -\tau_x\sin(\theta) - \tau_y\cos(\theta) + \tau_y \\ 0 & 0 & 1 \end{pmatrix} \quad (2)$$

where $\theta$ is the rotation angle, $\tau_x, \tau_y$ correspond to the distances to the rotation centre, and we define

$$t_x := -\tau_x\cos(\theta) + \tau_y\sin(\theta) + \tau_x \quad (3)$$

and

$$t_y := -\tau_x\sin(\theta) - \tau_y\cos(\theta) + \tau_y \quad (4)$$

the residual translations associated with a rotation around an arbitrary center $(c_x, c_y)$, which can be computed from a reference "zero center" (e.g., the center of the field of view $(c_{x_0}, c_{y_0}) := (M/2+1, N/2+1)$, where $M$ and $N$ represent the k-space/image matrix dimensions, as $c_x = c_{x_0} - \tau_x$ and $c_y = c_{y_0} - \tau_y$.

The rigid-body motion parameters associated with in-plane rotation around an arbitrary center are therefore $\theta$, $t_x$ and $t_y$.

Our hypothesis posits that, in general, motion-corrupted brain scans acquired with the classical Cartesian Spin-Echo technique can be corrected in a satisfying manner by estimating these three latter parameters, neglecting potential out-of-plane motion.

*B. k-space quality metric*

A quality metric, operating in k-space, is proposed. Motion-corrupted phase lines can be detected in k-space and a quality score can be computed. We noticed that when raw multi-coil k-space data is reconstructed with the ESPIRiT algorithm [31], phase encoding lines corresponding to motion onset undergo a loss of signal with respect to the adjacent ones (in the k-space magnitude). This phenomenon can be explained by the convolution theorem because coil sensitivity profiles (estimated by ESPIRiT) are point-wise multiplied in image space during reconstruction, resulting in a convolution with the coil profile's Fourier transform in k-space. This leads to signal loss near the discontinuities as shown in Figure 2a. The detection method is as follows. Considering each phase encoding line $k_y$ as a vector $\boldsymbol{x} \in \mathbb{C}^n$, we compute a modified version of its $p$-norm:

$$||\boldsymbol{x}||_p := -(\sum_{i=1}^{n} |(\log|x_i|)|^p)^{\frac{1}{p}} \quad (5)$$

Mathematically, if $0 < p < 1$, then Equation 5 defines a quasinorm rather than a norm, since it fails to satisfy the triangle inequality but retains the two other properties common to all norms (non-negativity and homogeneity). Classical signal processing approaches could tackle the problem by low-pass filtering the quasinorm 1D signal, and detecting peaks by thresholding discrete derivatives. However, a suitable threshold value is difficult to determine. An adaptive threshold (with respect to position in k-space) is also problematic, since we have observed that peak amplitude in the quasinorm signal (for a given angle) varies significantly, even for the same phase line. Hence a deep learning approach was chosen. It was cast as a multi-class classification problem (restricting the problem to the detection of a single motion event). A (fully) convolutional neural network (CNN) was trained to predict the index (class) of the PE line at which motion occurred, identified by a drop in signal at the corresponding line. After trying different CNN architectures (inception blocks, depths from 3 to 12 layers and regularization such as weight decay or dropout) the most economical and accurate model was a 3-layer CNN with 3, 6 and 9 convolutional filters respectively, with kernel sizes $1 \times 3$ and PReLU nonlinearities, for a total of 272 trainable parameters.

To improve detection and since Spin-Echo acquisitions are always multi-slice, we compute the quasinorm of the average of the magnitudes of all the slices and this (vector of 256 values) is the input provided to our classification CNN. Taking the average has the effect of emphasizing signal loss since all slices of the same acquisition experienced the same rigid motion as shown in Figure 2. A k-space quality score was then computed by multiplying the index of the detected PE line by a weight corresponding to its distance relative to DC.

This is based on the intuitive assumption that motion closer to DC has a higher impact on image quality. Finally, the metric was validated by three human experts, assessing image quality of both simulated and in vivo motion-corrupted acquisitions. Manual scoring was based on 4 perceptual motion artifact categories:

- Motion-free: the image shows no perceptible motion artifacts, although Gibbs ringing may be present. This does not interfere with the image's clarity or detail.
- Small: minor artifacts are perceptible but do not compromise the interpretation of the image.
- Medium: artifacts are clearly visible but the image may sometimes remain potentially usable.
- Large: severe artifacts are present, significantly degrading image quality and compromising clinical interpretation. Such images are typically discarded.

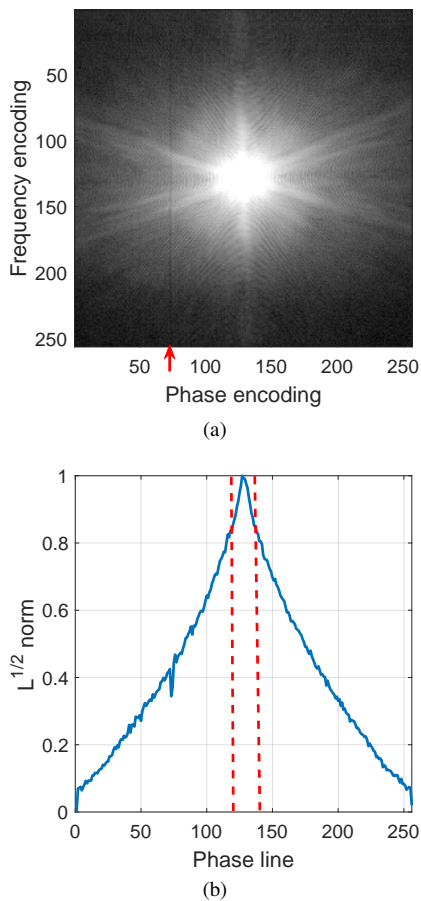

Fig. 2: The input to the line-detection CNN is the $L^{\frac{1}{2}}$ quasinorm of the logarithm of k-space magnitude (averaged over all slices). In (a) we can observe a visible loss of signal at phase line 75 (red arrow), where motion occurred and a corresponding peak in (b). The dashed red lines in (b) indicate excluded regions. (In-vivo k-space data from a volunteer who made a 3-degree rotation).

### C. DNN: Design, Objectives and Training Strategy

The proposed method leverages deep learning to quantify motion directly in k-space before applying a model-based reconstruction technique to remove motion artifacts. The assumption is that head movements are essentially rigid and in-plane. We considered only the phase encoding direction as the frequency encoding direction is sampled too fast for bulk motion to have any significant effects.

Our approach is based on the hypothesis that motion can be estimated from a limited number of adjacent lines in the phase encoding direction without any motion-free reference.

The proposed deep neural network (DNN) receives a block of consecutive k-space lines in the phase encoding direction as input, organized as a 2-channel tensor of real and imaginary components. The Fourier rotation theorem and the Fourier shift theorem establish corresponding relationships in image space and k-space. A rotation by $\theta$ radians in image space corresponds to a rotation by the same $\theta$ angle in k-space (around the DC point). Translations in image space correspond to phase ramps in k-space. The assumption is that the DNN will be able to learn these relationships and detect and quantify inconsistencies occurring due to a motion event.

Deep learning approaches require large amounts of data, it was hence necessary to simulate motion artefacts to build a training set of adequate size. To demonstrate the feasibility of learning and generalization in k-space without any reference, a deep convolutional neural network model was trained using single simulated motion events and subsequently tested on both individual events and more complex motion trajectories. Our hypothesis is that motion can be estimated in a *relative* manner (between temporally adjacent k-space lines). A complex motion trajectory can be approximated as a piece-wise constant (between PE lines) function and hence, training on single motion events should allow generalization to complex trajectories. Multiple instances of the same DNN model architecture were trained on different k-space regions (centered on a given PE line). Therefore, each DNN is in a way "specialized" for a given line (or more accurately, in a limited neighborhood of the given line).

The proposed DNN model architecture is similar to that of a LeNet [42] or AlexNet [43] convolutional feedforward artificial neural network. These networks can be conceptually separated into two parts: a feature-extractor followed by a classifier [44]. In our case, the second part is rather performing regressions since it has to output 3 real valued parameters: the rotation angle $\theta$ (degrees) and two translations $t_x$ and $t_y$ (pixels). The full network architecture is shown in Figure 3. It *searches* for *motion* parameters in *segments* of a full *k-space*, hence we called it SISMIK: Search In Segmented Motion Input (in) K-space.

*1) Input normalization:* For artificial neural networks to learn effectively, the relevant data points should all be in a homogeneous range. In order to cope with the high dynamic range due to the high DC component, we computed mean and standard deviation maps from all motion-free k-spaces in our motion-free in vivo dataset (used as a basis for simulation). During DNN training, each minibatch was then normalized by subtracting the mean and dividing by the standard deviation maps, point-wise. Following this normalization, the dynamic range of the k-spaces becomes suitable for deep learning.

*2) DNN training strategy:* Training of the DNN was carried out using supervised learning on simulated datasets. An input consisting of a small k-space window around motion onset was provided along with the ground truth motion vector of 3 real-valued parameters: $(\theta, t_x, t_y)$ for this single motion event. Therefore, the input consists of a motion-free region followed by a "transition" (typically one or two lines, depending on motion speed) and the final rotated region. Based on empirical results, a window size $w$ of 9 demonstrated superior performance and was selected for our experiments. We used the mean squared error (MSE) loss function as can be justified by the maximum likelihood principle for modelling Normally distributed errors [45]. During the training phase, the DNN learns the relationship between the corrupted k-space block (input) and the motion labels (output) in a supervised manner. It does *not* receive any motion-free reference (e.g., like a low-resolution scout scan for example) in its input.

SISMIK starts by extracting features at different levels through its convolutional layers (see Figure 3 "feature extractor block"). The first three $1 \times 3$ (padded) convolutions do not change the input size and allow a transformation of the input into a more suitable representation for further analysis. This can be thought of as the network performing an encoding of the $w$ phase lines into an internal representation, without merging them. Subsequently, a $w \times 3$ convolution collapses the feature map size into a shape of $256 \times 256$. This step allows the network to combine the information of all the lines in the input simultaneously. This is the penultimate part of the feature extractor block (see Figure 3) and allows the network to obtain high level features, before performing one more convolution step and a final regression of the motion parameters (performed by two dense layers). We may think of the first $1 \times 3$ convolutions as performing some kind of pre-processing of the input before a compression into a single channel which is subsequently downsampled by max-pooling and flattened to be fed to the final fully connected (dense) block. The latter performs the final regression of the 3 output rigid-body motion parameters. Training was performed with the PyTorch deep learning library using an Adam optimizer with an adaptive learning rate starting at $10^{-6}$, a batch size of 128, Tikhonov regularization at $10^{-6}$, a single 0.25 dropout layer and early stopping (based on the validation set performance). It is important to note that the initial learning rate of $10^{-6}$ had to be empirically chosen significantly lower than is usual for Adam (i.e., usually closer to $10^{-3}$), otherwise learning diverges. We used a total of 100 epochs for training with random shuffling of the trainingset each time. This was justified by the fact that no more improvements could be observed and the learning rate dropped below $10^{-8}$.

### D. Simulated and in-vivo datasets

Raw, motion-free, classical multi-slice T1w Spin-Echo k-space data for the training set was gathered from a clinical 3T MRI system (MAGNETOM Prisma fit, Siemens Healthineers, Erlangen, Germany) through a stringent manual inspection process. Out of 300 acquisitions, only 43 were deemed acceptable (motion-free) and included in the dataset. With an average of 30 slices per acquisition, this resulted in around 1290 slices in total, which were used as a basis for simulation. The number of coils per acquisition varied between 16 and 58, an in-plane resolution of 1 mm and slice thickness of 4 mm. Although, the matrix size and FOV may vary from scan to scan, all the data were resized to $256 \times 256$.

Simulations were performed on each coil individually by applying rotations in image space around a random center, empirically chosen corresponding to real-like motion. For each simulated PE line, the corresponding k-space rows in the motion-free slice were replaced by the corrupted k-space ones.

Around 600k training examples were simulated for each phase encoding (PE) line on which the DNN was trained. A validation set of 10k simulations and a simulated testset of 10k were also generated. The validation and test sets simulations were performed on 6 additional motion-free acquisitions unseen during training.

For demonstrating the network's capabilities in different k-space regions (spatial frequencies), the following PE lines were selected: 30,50,75,90 and 105. Motion parameters were drawn from a three-dimensional parameter space composed of the rotation angle and the center of rotation position. Parameters were independently drawn from a Gaussian distribution with mean $\mu = 0$ and $\sigma = 1.5$ for $\theta$ (degrees) and the rotation center offsets ($c_x$ and $c_y$) were uniformly drawn from the range $[-40, 40]$ (pixels) which resulted in $t_x$ and $t_y$ in (approximately) the range $[-2.5, 2.5]$ (pixels). Also, two different motion speeds were simulated: "instantaneous", i.e., abrupt transition from 0 to $\theta$ degrees and a "slower" speed with a small $1 \times 3$ convolution kernel around the transition.

Finally, an "in-vivo" testset was produced following the ChoCo protocol described in [46]. We asked 15 volunteers to perform different video-controlled head choreographies, to produce known motion artifacts on classical T1w Spin-Echo acquisitions, with similar parameters as those used for training. With TR=500 ms, TE=12 ms, TA=2 min. The FOV was roughly $220 \times 220$ mm, the voxel size of $1 \times 1 \times 4$ mm, a matrix size of $256 \times 256$ and 16 coils. After data quality checks (motion-free), only the data from 5 subjects was kept for data augmentation.

We generated "empirical labels" by motion estimation using rigid-body registration of a preliminary EPI training sequence, according to the proposed ChoCo motion protocol [46]. These labels were used for DNN performance evaluation.

### E. Model-based motion correction

A non-uniform Fourier matrix can be built from the knowledge of the rotational motion trajectory that occurred during acquisition. An (inverse) non-uniform Fourier transform (NUFFT) can then be applied on the corrupted k-space data to recover a motion-free complex image. We used Fessler's NUFFT toolbox for a very fast and memory-efficient implementation of this procedure in Matlab[40, 47]. An example trajectory for a single motion event can be seen in Figure 4.

Translational motion can be corrected by applying point-wise opposite phase ramps to the corrupted k-space lines. The motion correction process can be mathematically described by

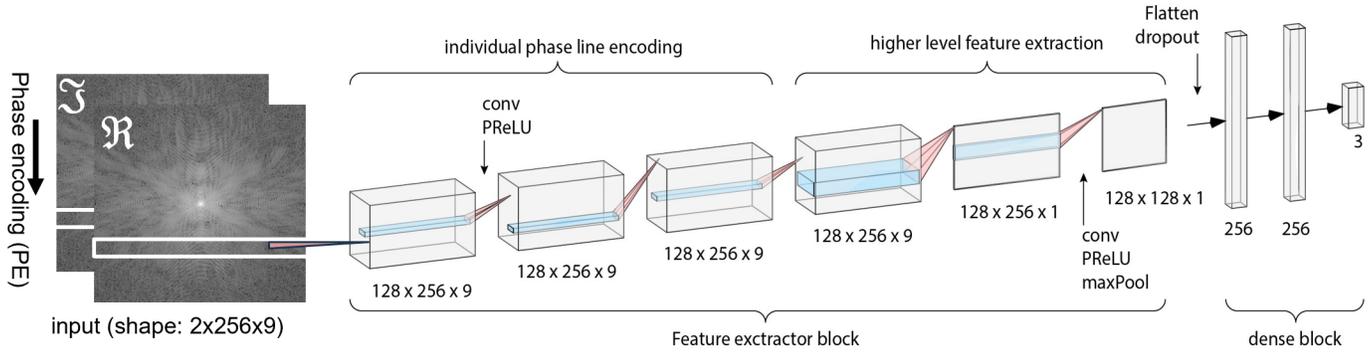

Fig. 3: SISMIK's architecture: 5 convolutional layers forming the "feature extractor" block, followed by max-pooling and flattening into a long vector of 16384 components (not shown), followed by a dense block for the final estimation of the 3 in-plane rigid-body motion parameters. The first 3 convolutional layers scan trough the feature maps, leaving their shapes unchanged until the penultimate convolution, which collapses it into a single channel. SISMIK has a total of 5 million trainable parameters.

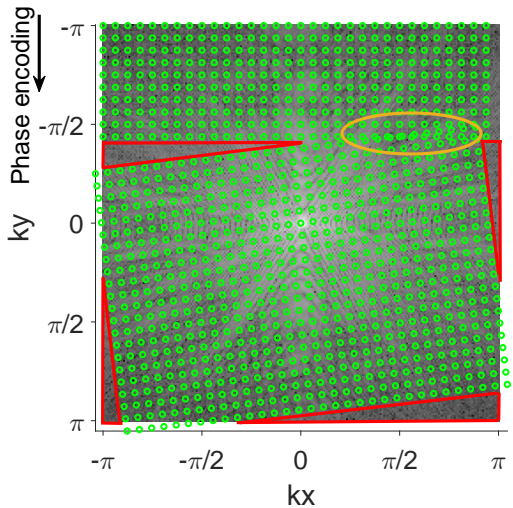

Fig. 4: Example of NUFFT trajectory for a single motion event occurring in the upper half of k-space. The grid (green dots) represents the theoretical simplified acquisition trajectory (undersampled) affected by one motion event. We can observe the missing k-space points in the "pie-slice" regions (in red) and an oversampled region (orange ellipse).

the following steps. Let $\Phi \in \mathbb{C}^{N \times N}$ the dephasing matrix be:

$$\Phi = e^{-i2\pi \boldsymbol{k}_x \odot \boldsymbol{t}_x} e^{-i2\pi \boldsymbol{k}_y^\top \odot \boldsymbol{t}_y^\top} \quad (6)$$

where $\odot$ is the Hadamard product, $\boldsymbol{k}_x$, $\boldsymbol{k}_y$ are vectors of spatial frequency coordinates and $\boldsymbol{t}_x$, $\boldsymbol{t}_y$ are vectors of estimated translations (for all spatial frequency coordinates, including the motion-free ones, for which the parameters are equal to zero).

$$\widetilde{\boldsymbol{Y}} = \boldsymbol{Y} \odot \Phi \quad (7)$$

where $\boldsymbol{Y} \in \mathbb{C}^{N \times N}$ is the measured motion-corrupted k-space matrix and $\widetilde{\boldsymbol{Y}} \in \mathbb{C}^{N \times N}$ the phase-corrected k-space.

$$\widehat{\boldsymbol{x}} = \boldsymbol{F}_{\text{NUFFT}}^\dagger \text{vec}(\widetilde{\boldsymbol{Y}}) \quad (8)$$

where $\widehat{\boldsymbol{x}} \in \mathbb{C}^{N^2}$ is the reconstructed (vectorized) image, the dagger (†) denotes the Hermitian adjoint of $\boldsymbol{F}_{\text{NUFFT}}$, the non-uniform Fourier transform operator and $\text{vec}(\widetilde{\boldsymbol{Y}}) \in \mathbb{C}^{N^2}$ is the notation for a vectorized matrix, i.e., $\widetilde{\boldsymbol{Y}} \in \mathbb{C}^{N \times N}$ has all its columns flattened into a long vector of $N^2$ components.

In practice, better results, with no significant computational cost, are obtained by iteratively computing:

$$\widehat{\boldsymbol{x}}_{\text{LSQR}} = \text{argmin}_{\boldsymbol{x}} ||\boldsymbol{F}_{\text{NUFFT}} \boldsymbol{x} - \text{vec}(\widetilde{\boldsymbol{Y}})||_2^2 \quad (9)$$

for about 10 or 12 iterations (it takes less than a second), which we will refer to as the iterative least-squares NUFFT.

Image reconstruction using inverse NUFFT was performed on the ESPIRiT multi-coil combined k-space.

### F. Experiments and performance evaluation

Dataset simulations were performed on a HPC cluster and neural network training was done on a mid-range server equipped with a CUDA-compatible GPU with 12GB of VRAM (NVIDIA TITAN V).

Training takes on the order of a day for a 600k dataset on a single GPU and inference times are on the order of the second for a batch of ($\sim 25$) slices. Inference times for a single slice using the GradMC method (described in subsection II-F.3) are on the order of 1-2 minutes.

*1) k-space quality metric optimization:* To train our corrupted-line-detection CNN, we simulated single motion events at random positions in k-space in 240k full multi-slice acquisitions. From the latter, the $p$-quasinorms of the average k-space magnitudes were computed. Therefore, we needed to first identify a $p$ quasinorm value that would enhance peak visibility. This was estimated from simulations calibrated to cover the range of PE lines and a discretization of angle magnitudes to cover a clinically-relevant range (between -4 and 4 degrees). The PE lines were chosen between 10 and 250 by steps of 5 with 16 lines removed around the DC line (dashed region in Figure 2). The chosen angles were $\theta \in \{0, 0.01, 0.1, 0.2, 0.3, 0.4, 0.5, 1.0, 1.5, ..., 3.5, 4.0\}$. We considered $\theta < 0.3$ as "no motion". This choice was dictated by the fact that motion artifacts are barely visible at $\theta \simeq 0.5$ degrees and, in general, not visible at all to the human eye below that. With two different speeds (same as for training) and 10 different motion-free acquisitions used for simulating the artifacts, this resulted in a total of 5244 (full multi-slice volume) simulations.

ROC curves were plotted to give an estimation of the trade-off between the true positive rate and the false positive rate for different values of $p$ according to Equation 5. This allowed to empirically determine the best $p$ norm value.

Validation of the k-space quality metric was performed by three experts who rated a set of 960 simulated acquisitions covering half of k-space (corrupting the other half, being conjugate-symmetric, would generate visually similar artifacts). A set of 85 in vivo acquisitions (single motion events at PE lines 50, 75 or 90 and motion-free) was also rated by the experts.

*2) Motion estimation accuracy:* Motion parameter estimation accuracy was evaluated on in-silico test sets (with motion simulations performed on motion-free acquisitions that were unseen during training) using the RMSE and Pearson correlation coefficients between the ground truth and the network's predictions. For in-vivo test sets, we computed the median of SISMIK's predictions on all slices of each acquisition and this was used as the estimated label for the corresponding acquisition (since all slices in the same acquisition undergo the same rigid motion). The RMSE was reported for each case.

*3) Motion correction quality:* Next in our approach, a comparison of the reconstruction quality with respect to an auto-focusing approach is proposed. The freely available GradMC method was chosen [21] because it uses a model-based correction approach, similar to ours, based on a matrix inversion formalism using an entropy metric [18, 48] to assess image quality. PSNR and SSIM metrics were computed from motion-free ground truth and motion-corrected simulations for both strategies.

In-vivo motion correction quality was assessed using information gain as a quantitative metric. Information gain can be defined as the difference between the Shannon entropy of a corrupted image and the corresponding restored image. An improvement in image quality should decrease the entropy, resulting in positive information gain. We also provide representative examples of in-vivo corrected images for qualitative reconstruction assessment.

GradMC was run for 100 iterations for each testset case. This number of iterations was chosen because no significant drop in entropy occurred afterwards.

*4) Additional motion correction evaluation:* The approach was further tested on multiple motion events and different contrasts. The SISMIK model, trained exclusively on T1-weighted (pre/post-Gd) acquisitions and single motion events, was applied to estimate motion in a T2-weighted Spin-Echo acquisition. This application, as illustrated in Figure 11, resulted in a significant reduction of motion artifacts. Additionally, Figure 11 demonstrates the model's ability to accurately estimate motion for multiple motion events in both T2-weighted and T1-weighted motion-corrupted acquisitions obtained by simulation. Furthermore, Figure 12 shows that SISMIK successfully estimated in vivo motion for multiple motion events.

### G. Inference

Multiple instances of the same SISMIK DNN architecture were trained, each specialized for a given PE line. Ideally one instance $SISMIK_i$ is trained for each PE line $i$ but $SISMIK_i$ performs reasonably well in a limited region of a few PE lines around the $i$-th line (on which it was trained). Note that no *a priori* knowledge of the corrupted regions in k-space is required. Given $n$ instances of trained SISMIK networks, motion parameters of each of the $n$ lines of k-space can be estimated in parallel. Subsequently, a motion trajectory can be constructed from these $n$ estimations.

## III. EXPERIMENTAL RESULTS

*1) k-space quality metric:* The optimal $L_p$-quasinorm value was selected from ROC (Receiver Operator Characteristic) curves assessing detection (motion vs no motion) performance in terms of true positive rate and false positive rate.

Results are presented in Figure 5 for 5244 motion simulations on a range of PE lines covering the whole k-space expect for a small region near the center (DC) as described in subsubsection II-F.1. ROC curves for 7 different $p$-quasinorm values were computed for 1000 different thresholds and are presented in (Figure 5). A true positive rate close to 90% can be obtained for a false positive rate near 4% by choosing $L_p$ with $\frac{1}{16} \leq p \leq \frac{1}{2}$. The corresponding AUC is 95%. Higher order norms ($p \geq 1$) have a strictly worse performance.

The corrupted line detection neural network achieves 5% error rate from $L_{1/2}$ quasinorms.

Results in Table I and Figure 5(b,c) validate the k-space quality metric with respect to manual expert perceptual scoring on 960 motion-simulated acquisitions (b) and 85 in vivo motion-controlled acquisitions (c).

*2) Motion estimation with SISMIK:* Motion parameter estimations obtained with SISMIK are presented for increasing wavenumbers and the RMSE and Pearson correlation metrics were used to provide two complementary views of the SISMIK's performance. RMSE and Pearson correlation results are shown in Figure 6. Performance increases with decreasing spatial frequency, as expected following the SNR distribution in k-space acquisitions. The relationship between SNR level and inference accuracy is not explored in this paper (see [49, 50] for general considerations). Translations ($t_x$, $t_y$) exhibit lower correlation than the rotation angles and seem more difficult to learn, the correlation metric being independent of error magnitude. For an intermediate spatial frequency (PE=75), SISMIK achieves an RMSE around 0.55 degrees for the rotation angle and an RMSE around 0.35 pixels for the translations. Rotation angle estimation results obtained on simulated testsets ($\sim$ 10k samples) can be seen in Figure 7. A rolling window of 100 data points was used to display the mean and standard deviation. SISMIK starts to exhibit a larger standard deviation at the extremities of the range on which it was trained but otherwise shows good performance.

For in-vivo testsets acquired with the ChoCo protocol [46], performance is as follows. Experiments were performed with volunteers on PE=75. The RMSE of SISMIK estimations compared to ChoCo labels was 0.59 degrees for the rotation angle and 1.21 pixels for $t_y$ and 0.18 pixels for $t_x$. The in-vivo angle RMSE is very close to the in-silico RMSE (0.55). The $t_y$ RMSE is higher due to the presence of two significant outliers, which correspond to the large negative information gains for subject 4 in Figure 10.

*3) Motion correction from SISMIK predictions:* The last stage of the pipeline entails reversing the effects of motion using a model-based approach by building a NUFFT matrix from the estimated rotation angles following phase ramp cancellation with the formula in Equation 7. Qualitative results for simulations are shown in 8a and are also compared to the GradMC method. Quantitative results presented as boxplots for a range of angles and PE lines covering different motion intensities and k-space regions are shown in 8b and 8c. NUFFT reconstructions with SISMIK estimations result in a median PSNR of 37.8 dB and a median SSIM of 0.98. For the same slices corrected with GradMC, a median PSNR of 29.7 dB and SSIM of 0.93 are obtained.

By only focusing on the "Corrupted" and "SISMIK" boxplots in Figure 8 (b) and (c), we can compare the PSNR and SSIM metrics of corrupted and restored slices, solely from SISMIK estimations, with respect to the motion-free ground truth used for simulation. All the restored slices obtain a PSNR > 30 dB, whereas all the corrupted ones are below 30 dB. For the SSIM metric, all corrupted slices are below 0.91 and all the restored ones (except for a few outliers) are above 0.97. Overall, this demonstrates that the proposed DNN estimations allow a model-based restoration with NUFFT that can completely separate the motion-corrupted PSNR and SSIM distributions from the restored ones.

Two representative examples of in-vivo motion-corrupted ChoCo slices and their restored versions based on SISMIK predictions are presented in Figure 9. These qualitative results show that SISMIK estimations were accurate enough to allow the removal of most of the motion artifacts. Moreover, Figure 11 presents a representative example of multiple simulated motion events with two different contrasts (T1w and T2w) and corresponding restored brain slices after motion correction with the proposed approach. Figure 12 shows a similar example of an in vivo corrupted brain slice, restored from SISMIK estimations and model-based reconstruction.

Quantitative results for 5 in-vivo ChoCo subjects are shown in Figure 10. The latter distributions show that SISMIK estimations tend to get worse with increasing angle magnitude. It also shows that all (except 3 outliers) of the smaller angles (around 2 degrees) were well estimated. It can also be observed that 80% of the subjects (4/5) exhibit positive information gain for more than 75% of all their restored acquisitions. All the acquisitions (small, medium and large angles) of subject 2 were restored with positive information gain. Subject 4 represents a failure of SISMIK to estimate larger angles accurately.

| Class | Sensitivity | Specificity |
|---|---|---|
| Motion-free | 0.90 | 0.97 |
| Small | 0.89 | 0.99 |
| Medium | 0.97 | 0.80 |
| Large | 0.97 | N/A |

TABLE I: Sensitivity and specificity for manually assigned perceptual motion classes. The reported data was computed from the same 960 simulated acquisitions used to compute distributions in Figure 5(b). For the large motion artifacts class, no true negatives or false positives were found, resulting in undefined specificity (N/A).

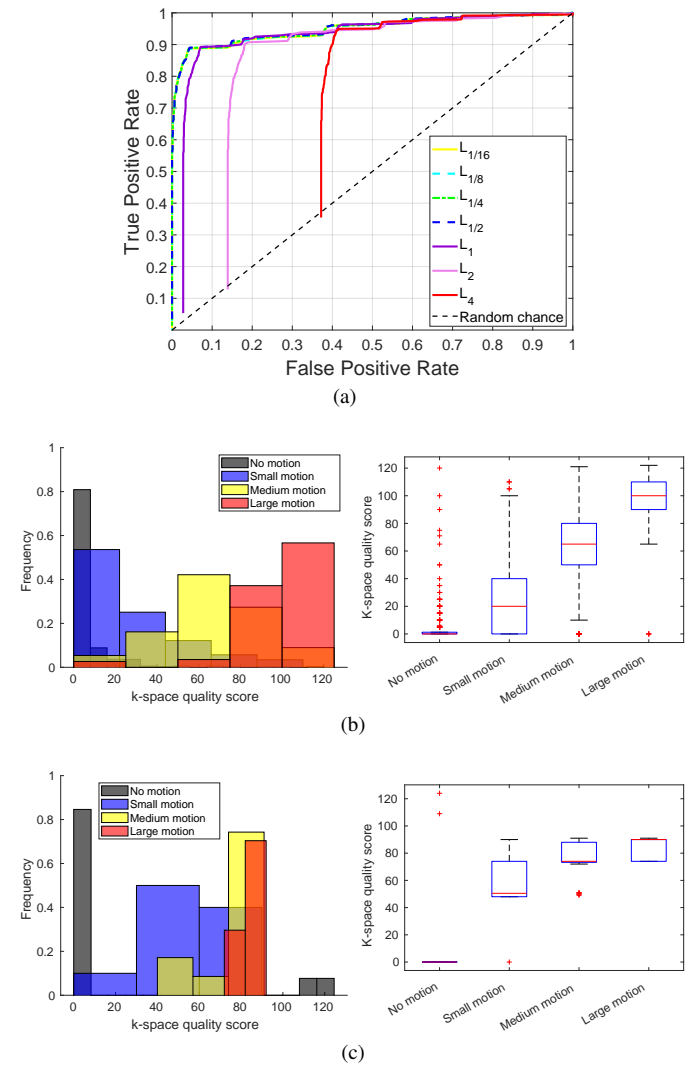

Fig. 5: (a) ROC curves for 7 different $L_p$ values and 1000 different thresholds. For $p < 1$, curves align closely, indicating that they provide similar sensitivity and specificity. A value of $0.3°$ was defined as the "empirical zero", below which any motion is deemed undetectable. (b) Distributions of 960 simulated motion-corrupted acquisitions classified by 3 human experts. (c) same as (b) but performed on 85 in vivo acquisitions. (See discussion for details about the two outliers in (c)).

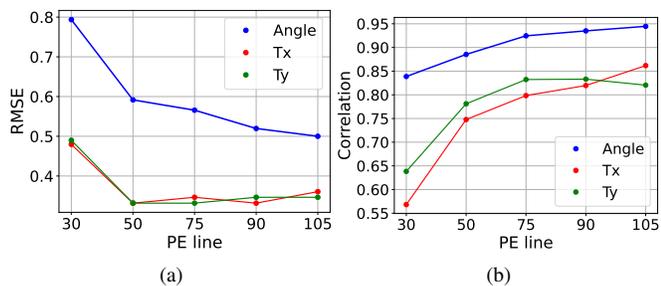

Fig. 6: DNN training performance (RMSE and Pearson correlation) for the estimation of in-plane rigid-body motion parameters. The x-axis goes from the highest (PE=30) spatial frequency to the lowest (PE=105) used in our experiments. Performance increases as we move towards lower spatial frequencies. (DC at PE=129).

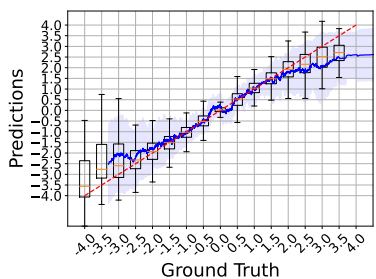

(a) PE = 30

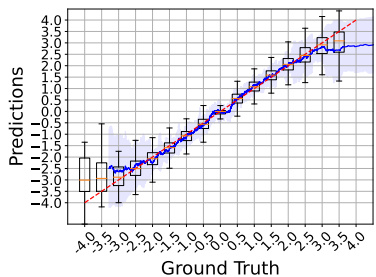

(b) PE = 75

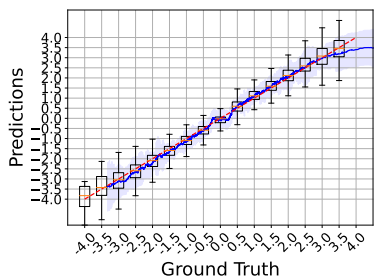

(c) PE = 105

Fig. 7: DNN estimations of the rotation angle for 3 different k-space regions: high (PE=30), medium (PE=75) and lower (PE=105) spatial frequencies. The latter obtains the best performance. The rolling mean and standard deviation (rolling window of 100 points) are shown in dark blue and as the shaded regions around it, respectively.

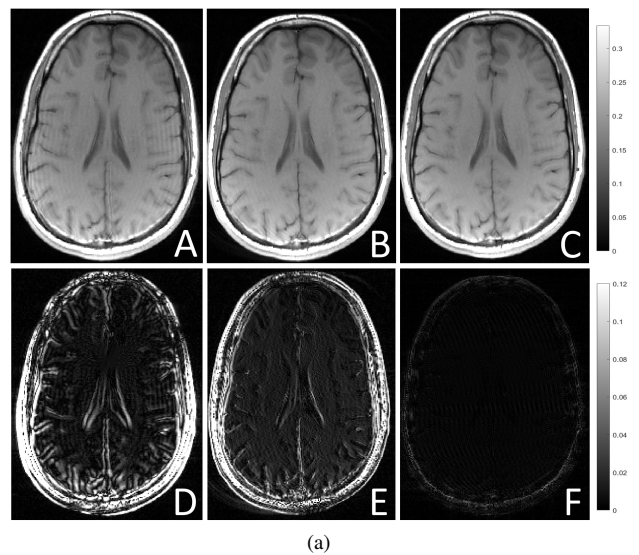

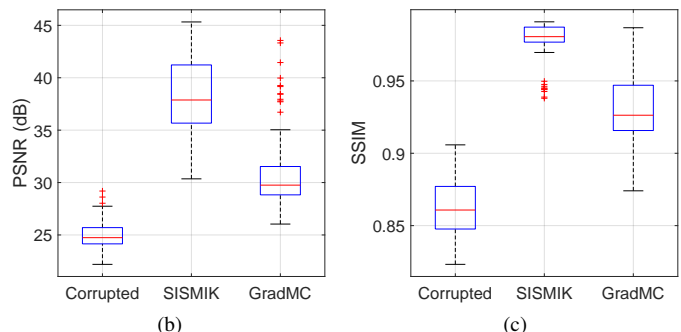

Fig. 8: (A) Representative example of simulated corruption (approx. 3 degrees rotation). (B) Restored with GradMC. (C) Restored with NUFFT from SISMIK predictions. (D,E,F) are the absolute difference to the reference motion-free acquisition used for simulation. PSNR and SSIM distributions for simulations over a range of [-4°, 4°] for different subjects and slices as shown In boxplots (b) and (c). In (b) the performance of SISMIK is compared to the GradMC method. In (c) we can observe that the distributions of PSNRs and SSIMs of the corrupted and restored (both with SISMIK) slices, do not overlap.

## IV. DISCUSSION

Experimental results demonstrate that it is possible to learn rigid-body motion parameters in k-space without a motion-free reference even in the higher spatial frequencies.

In order to evaluate the performance of SISMIK in high spatial frequencies, the system was trained on 5 different PE lines. The assumption being that as we deviate further from DC, the noise level increases, thereby complicating the learning process. This assumption is verified, but generalization performance in higher spatial frequencies - which were excluded in previous work [17] - is still sufficient to allow for qualitatively and quantitatively acceptable motion correction quality, as shown in Figure 6. Work by Rajwade *et al.* in [15], started to explore motion estimation from two adjacent k-space lines using a convolutional neural network. Their experiments remained *in-silico*, only leveraging k-space magnitude to predict rotational motion. This preliminary work inspired us towards the exploration of full rigid-body motion estimation in k-space.

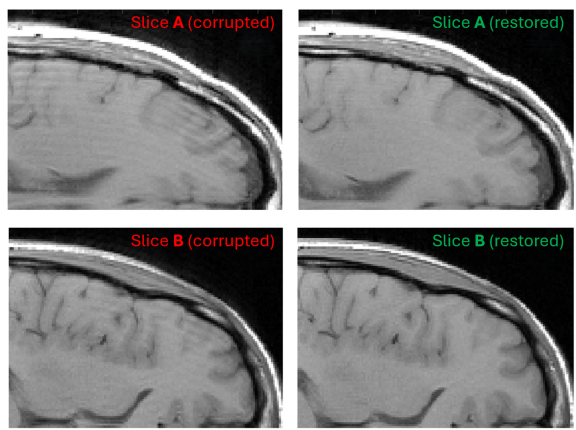

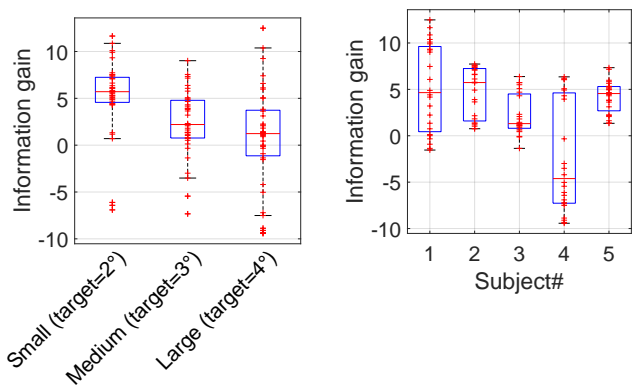

Fig. 9: (Left column) Representative examples of classical T1w Spin-Echo acquisitions corrupted in-vivo following the ChoCo procedure for generating controlled motion artifacts. (Right column) Restored images by phase cancellation and iterative NUFFT reconstruction from DNN motion parameter estimations. The two rows show the same anatomical region for two different subjects.

Fig. 10: (Left) information gain distribution for 3 (in-vivo ChoCo) angle categories corresponding (roughly) to 2, 3 and 4 degrees. Performance decreases with increasing angle magnitude. (Right) Information gain distributions for 5 ChoCo in-vivo subjects. All have more than 75% positive information gain, except subject 4.

Restoring high spatial frequencies is especially important to preserve fine details provided by state-of-the-art MRI technology. Note that all our datasets were resized to $256 \times 256$ (from matrix sizes ranging from 234 up to 270 phase encoding lines and 256 or 320 frequency encoding lines). For smaller matrix sizes, we simply performed zero-padding. For a k-space matrix larger than $256 \times 256$, the matrix should be cropped. However, this has the disadvantage of ignoring high spatial frequencies. To obtain the most general model, source data with a larger matrix size (e.g., 512 or more) could be used to generate the training set, hence including all lower resolutions. Please, note that we did not observe any impact on performance with padded k-spaces.

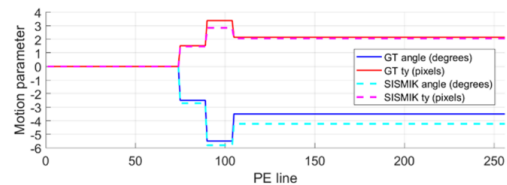

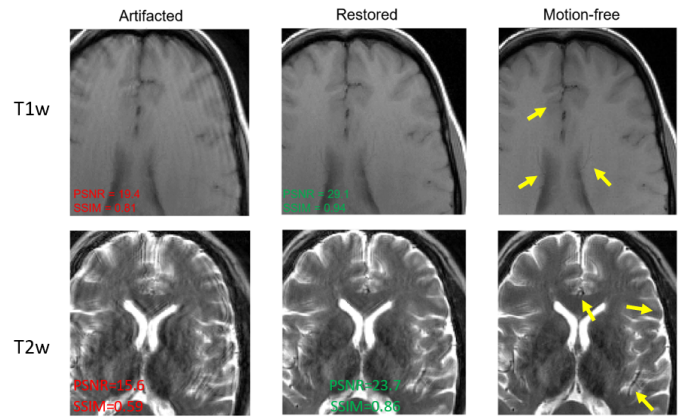

Fig. 11: Simulated motion trajectory with 3 movements in T1w and T2w Spin-Echo acquisitions. Solid lines indicate the ground truth (GT) and the dashed lines show SISMIK estimations. Yellow arrows indicate fine structures in the motion-free image that were obscured by motion and recovered after restoration with our approach.

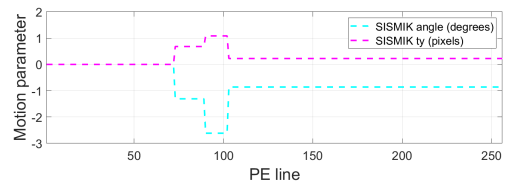

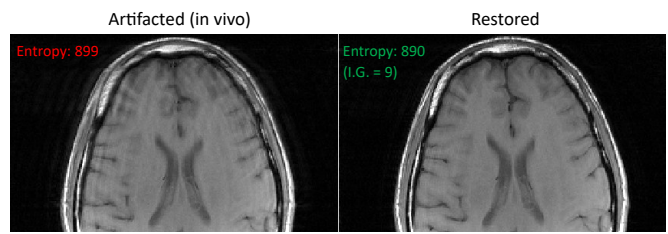

Fig. 12: SISMIK's estimation of an in vivo motion trajectory with 3 motion events performed by a volunteer (above). Motion corrupted and restored images. Information gain (IG) is significantly increased (9 points). Qualitatively, we observe a significant reduction of blurring and ringing artifacts.

SISMIK is designed for Cartesian k-space sampling but could be adapted to non-Cartesian trajectories (e.g., spiral, radial, etc.), thanks to the sequential acquisition process of MRI. The non-Cartesian k-space data would need to be re-arranged into rows (time series) that correspond to the temporal order of the acquisition, before feeding to the CNN. Different architectural parameters such as the number of layers and the number and placement of dropout modules were investigated before arriving at the proposed model shown in Figure 3.

Larger rotation angles are more difficult to learn due to the limited standard deviation of the simulation distribution used, as well as the increasing loss of information at the edges of k-space (a.k.a. "pie slice" phenomenon) [51, 52]. This loss of information results in residual artifacts, even when reconstructing a slice using known simulation parameters. To enhance the accuracy of estimating large angles, it is necessary to increase the standard deviation of the Gaussian distribution for rotational angles. This adjustment will ensure that the training set includes more instances of high-amplitude motions. Additionally, a substantially larger training set is required to provide the deep learning model with sufficient examples of each angle, as suggested by preliminary findings (not included in this paper). Therefore, there is a trade-off between extending the motion range and increasing the size and accuracy of the training set. However, achieving motion correction for very large angles (e.g., 5 degrees or more) remains unlikely without significant residual artifacts, which could be problematic.

A significant challenge was to properly normalize k-space to enable learning. Different normalization methods were investigated like the classical minibatch standardization or using global dataset statistics (mean and standard deviation). We observed that learning was only successful by normalizing with mean and standard deviation maps computed from the motion-free dataset ($\sim$1000 slices from 43 acquisitions). This resulted in a more homogeneous k-space allowing deep learning to process features roughly on the same scale.

A k-space quality metric is also proposed. It leverages the fact that multi-coil ESPIRiT reconstruction enhances motion transitions in k-space, which are detectable using quasinorms and discrete derivatives, followed by thresholding. Finding a single or even multiple thresholds revealed challenging, and much better results were obtained by training a small fully convolutional neural network for corrupted line detection. For Spin-Echo acquisitions, it is possible to leverage the Hermitian symmetry of k-space to detect signal drops near the corresponding motion affected PE lines[53, 54]. This approach works to some extent but Hermitian symmetry is not good enough in practice to yield better results than the proposed method. This quality metric serves as a valuable tool during the dataset generation process. It offers a convenient method for performing "sanity checks" to ensure that simulated or in-vivo motion events are correctly localized in k-space. Moreover, expert scoring is in good agreement with the k-space quality metric as shown in Figure 5(b, c). In the motion-free class distribution in Figure 5(c), we observe two outliers that were assigned a high k-space degradation score. These are false positives generated by a mistake of the corrupted line classification neural network. When the corresponding k-space lines ($109 \pm 5$) are presented to SISMIK, it predicts very small rotations (approx. $0° \pm 0.15°$) and similarly small translations, indicating that no motion was detected. Visual inspection of the $L_{1/2}$ quasinorms of the average k-space reveals the presence of a small peak that has most likely misled the k-space quality metric network. The peak appears symmetrically on both sides of DC, indicating that it is inherent to the object's structure and most likely not a motion artifact. The k-space quality metric approach is very sensitive to medium and large motion artifacts as can be seen in Table I. Note that the specificity for the class of large motion artifacts is undefined, as there were neither true negatives nor false positives for this class. The proposed k-space quality metric could also be a good complement or alternative to image-based metrics. A limitation of the approach is that k-space magnitude is is averaged over all slices of an acquisition, which assumes that they were corrupted by the same motion (which is the case for classical Spin-Echo). To overcome this limitation, the method could be adapted to only consider slices that were acquired quasi-simultaneously.

Working in k-space confers the advantage of precise motion localization, in contrast to image space where it is distributed everywhere. SISMIK is able to leverage local information with a narrow window of only 9 PE lines to estimate motion parameters. Note that we experimented window sizes from 3 up to 17 and it was empirically observed that 9 exhibited optimal performance (data not shown). Smaller sizes likely lack sufficient information, while larger sizes may overwhelm the network with excessive data. This comes at the cost of having multiple DNNs specialized for each phase encoding line using the same architecture, just trained on a different k-space block. In practice we have observed that the networks are able to obtain a similar performance for PE lines in a close neighbourhood (around $\pm 5$ lines). Therefore, a potential five-fold reduction in the number of trained models may be possible, but this requires further investigation and testing. However, once training is complete, all the phase encoding lines of a motion-corrupted k-space slice may be passed in parallel, hence incurring no cost related to the "multi-network" requirement.

It is a well-known fact that deep neural networks are prone to "hallucinations", e.g., the creation of spurious structures in a brain image, which can lead to diagnostic errors [11, 55]. Therefore, we used a model-based reconstruction approach rather than an "image to image" strategy such as [6] or [4], embedding a full reconstruction framework inside a DNN. In the worse case, the model-based approach will fail to remove enough artefacts but it will not introduce false structures or remove existing ones. When the motion-estimating-DNN is able to provide the correct parameters and the in-plane rigid-body assumption holds, then artifacts may be significantly removed.

A comparison with the GradMC model-based method [21] was performed. The latter uses an "autofocusing" approach based on the optimization of an entropy metric to estimate motion parameters of a corrupted MRI acquisition. Figure 8 shows that SISMIK's estimations of the motion parameters followed by phase cancellation and iterative NUFFT achieves a higher increase in PSNR and SSIM. GradMC is also less robust in the sense that it can often diverge and is unable to find the optimal motion parameters. For this comparison, GradMC was initialized with motion parameters obtained from SISMIK, in order to avoid a divergent behaviour. This highlights the fact that such autofocusing approaches can benefit from better initialization. SISMIK can provide these estimations and therefore reduce the computational burden of such iterative methods, which may require in the order of minutes to correct a single slice.

Overall, SISMIK is able to restore motion-corrupted slices simulated from a parameter space that corresponds to realistic patient motion. In practice, with properly padded head-coils, only small rotation angles are expected[56, 57].

Figure 8 also shows that the PSNR and SSIM distributions of motion corrupted and restored images are consistent and do not overlap after motion estimation and reconstruction with the proposed approach, showing improvement in all simulations.

Figure 9 shows representative examples of corrupted slices along with the restored versions, which qualitatively highlights the ability of SISMIK to correctly estimate *in-vivo* motion parameters. Small angles are better estimated than larger ones, resulting in better image quality (higher information gain) for angles $\leq 2$ degrees per motion event, in general (Figure 10 (left) ). Information gain distributions for all in-vivo subjects - acquired with the ChoCo protocol - and all angles (2,3 and 4 degrees for each subject) are presented in Figure 10 (right). All subjects show more than 75% information gain, except subject 4. This indicates that SISMIK has successfully estimated motion parameters in-vivo for the vast majority of cases, with failures predominantly isolated to a single subject. It may be related to the fact that this subject has performed larger angles than the others during the ChoCo in-vivo acquisitions. It is noteworthy to mention that our approach is able to recover fine structures that were obscured by complex motion trajectories (multiple motion events) and are likely to hinder a radiologist's diagnosis, as shown in Figure 11 and Figure 12. This demonstrates the ability of SISMIK to estimate motion parameters in a *relative* manner and construct the resulting complete motion trajectory. It is noteworthy that SISMIK was trained on single motion corrupted k-spaces and was able to generalize to multiple motion events and T2 contrast in the same acquisition.

Different directions are possible for improving the approach. It is possible to train the same DNN architecture with simulations performed on individual coils rather than multiple coils combined with ESPIRiT, as was performed for this proof of concept. The signal drops in k-space due to convolution with the coil sensitivity profile's Fourier transform will no longer occur. We will need to investigate whether this change enhances or hampers the learning process. Another possibility involves adapting the method to 3D acquisitions. This extension should be achievable as long as the k-space data is presented to the network in the order of its sequential acquisition. This requirement exists because the DNN employs local temporal motion information to estimate the motion trajectory. In particular, this could allow restoring motion corrupted 3D MP-RAGE acquisitions, which are widely used in the clinical setting.

In the future, it would be interesting to investigate complex-valued neural networks (CVNNs), which directly handle complex numbers, efficiently halving the number of parameters by ensuring real and imaginary components are not treated independently, thereby improving statistical efficiency [58]. Despite their theoretical advantages, CVNNs remain marginally supported by deep learning frameworks and may not be well optimized yet, limiting their practical application. Nevertheless, implementing a CVNN version of SISMIK could be a promising future work to investigate.

As another future work direction, we hypothesize that incorporating "frequency attention layers" into a new SISMIK variant, taking a full k-space as input, could allow the model to more effectively focus on relevant spatial frequencies [59, 60]. This approach aims to prevent the model from being overwhelmed by the sheer volume of data in the full k-space and possibly discard irrelevant information.

Following a new trend, MRI scanner vendors now offer tools for running reconstruction or post-processing algorithms directly on the scanner. We are considering integrating our SISMIK pyTorch scripts with these tools. This integration would provide motion-corrected images that radiologists can quickly access after an MRI scan. They could then decide whether to keep the images or perform a re-scan. To achieve this goal, further refinements, and optimizations of the SISMIK approach would be required like designing a global SISMIK network capable of receiving a full k-space as input and further validating motion estimation on random clinical scans.

## V. CONCLUSION

In this study, a novel deep learning-based approach for quantifying in-plane rigid-body motion in k-space was proposed. It is novel in at least two aspects: it is reference-less and is able to learn in the higher k-space spatial frequencies, despite an increasing amount of noise. A novel k-space quality metric was also proposed, which leverages signal drops arising from convolutions in k-space when ESPIRiT reconstruction is performed. The experimental results demonstrate that the proposed DNN is able to obtain good generalization performance for motion estimation followed by correction of corrupted images obtained in-silico as well as in-vivo.